\documentclass[aps,reprint,pre]{revtex4-2}
\usepackage{graphicx}
\usepackage{amsmath}
\usepackage[hidelinks]{hyperref}
\usepackage{multirow}

\begin{document}

\title{BEER: Biochemical Estimator \& Explorer of Residues\\
A Comprehensive Software Suite for Protein Sequence Analysis}

\author{Saumyak Mukherjee}
\email{mukherjee.saumyak50@gmail.com; saumyak.mukherjee@biophys.mpg.de}
\affiliation{Department of Theoretical Biophysics, Max Planck Institute of Biophysics, 60438 Frankfurt am Main, Germany}

\begin{abstract}
Protein sequence analysis underpins research in biophysics, computational biology, and bioinformatics. We introduce \texttt{BEER}, a cross‐platform graphical interface that accepts FASTA or Protein Data Bank (PDB) files—or manual sequence entry—and instantly computes a suite of physicochemical metrics: amino acid composition; Kyte–Doolittle hydrophobicity profiles; net charge versus pH curves with automatic isoelectric point determination; solubility predictions; and key indices such as molecular weight, extinction coefficient, GRAVY (grand average of hydropathicity) score, instability index, and aromaticity. \texttt{BEER}’s interactive visualizations—including bar and pie charts, sliding‐window plots, residue‐level bead models, and radar diagrams—make it easy to explore physico-chemical properties of protein chains. A multichain module also enables direct comparison of complex assemblies. Built in Python with BioPython, PyQt5, and matplotlib, \texttt{BEER} delivers complete analyses of sequences up to 10\,000 residues in under one second.
\end{abstract}

\maketitle

\section{Introduction}

Proteins are essential biomolecules whose functions—from catalysis to structural support—derive directly from their amino acid (AA) sequences.  The order and frequency of AA residues govern folding, stability, interaction interfaces, and ultimately biological activity.  Although modern tools such as AlphaFold can predict three-dimensional structures from sequence alone, many early decisions in protein research—choice of expression system, buffer formulation, or mutagenesis targets—benefit from a rapid, sequence-based overview of key chemical features.  A quick assessment of residue composition, charge distribution, and hydrophobic or hydrophilic segments builds the chemical intuition needed to design experiments and interpret results long before detailed structural or thermodynamic studies are undertaken.

To address this need, we have developed \texttt{BEER} (Biochemical Estimator \& Explorer of Residues), a standalone, Python-based application with a clean, intuitive graphical user interface.  \texttt{BEER} integrates well-validated algorithms—amino acid composition analysis; sliding-window hydrophobicity profiling using the Kyte–Doolittle scale \cite{KyteDoolittle1982}; net charge and isoelectric point calculation via the Henderson–Hasselbalch equations \cite{Henderson1908}; and a suite of physicochemical indices (molecular weight, 280 nm extinction coefficient, GRAVY score, instability index, and aromaticity \cite{Guruprasad1990}), together with a simple GRAVY-based solubility heuristic—into a single desktop tool.  Results are presented through interactive bar and pie charts, sliding-window curves, charge–pH plots, residue-level bead models, and radar diagrams, and a multichain module enables direct comparison of complex assemblies from FASTA or PDB inputs. By combining comprehensive analytics with visually intuitive displays, \texttt{BEER} empowers students and researchers to obtain a fast, clear snapshot of any protein sequence—up to 10\,000 residues—in under a second, guiding experimental design and hypothesis generation from the very first glance. In the following sections, we describe \texttt{BEER}’s features in detail, explain the theory behind each calculation, and present examples illustrating how this tool can streamline both teaching and research workflows.

\section{Methods and Implementation}

\subsection{Software Architecture}

\texttt{BEER} adopts a Model–View–Controller design. The application is written in \texttt{Python} and builds on three core libraries: \texttt{BioPython} for all biochemical calculations \cite{Cock2009}, \texttt{PyQt5} for the graphical interface \cite{PyQt5}, and \texttt{matplotlib} for data visualisation \cite{Hunter2007}. Users can supply input in three ways—by importing a \texttt{FASTA} file, importing a \texttt{PDB} file, or directly pasting a single-letter amino-acid sequence. Only the 20 standard amino acids (\texttt{ACDEFGHIKLMNPQRSTVWY}) are accepted; any invalid characters trigger a warning dialog. Once a sequence is loaded, \texttt{BEER} automatically runs the analyses described in the following subsections. Note that if the input is provided by pasting the sequence, the user has to click the \texttt{Analyze} button for the analysis to take place.

\subsection{Amino Acid Composition} 
\texttt{BEER} counts the number of times each one-letter amino-acid codes appears in the sequence. The counts
$n_i$ are obtained with \texttt{Bio.SeqUtils.ProtParam} and the program converts them to percentage frequencies (Eq.~\ref{eq:count}).

\begin{equation}
    f_i = 100 \times \frac{n_i}{\sum_{j=1}^{20} n_j}.
    \label{eq:count}
\end{equation}

Alongside the numbers, \textsc{BEER} plots two graphs: a) a bar chart with each bar labelled by its percentage, and b) a pie chart that shows the relative proportions. Pointing at a slice reveals the exact count and percentage.

Seeing the composition at a glance helps users judge overall charge or hydrophobic character before running heavier calculations, spot unusual residue enrichments when comparing related proteins, and choose regions for truncation or mutagenesis based on which residues dominate the profile.

\subsection{Hydrophobicity Profiling} 

\begin{table}[ht]
  \centering
  \caption{Kyte–Doolittle hydrophobicity scores for the 20 standard amino acids.}
  \label{tab:KD}
  \begin{tabular}{lc}
    \hline
    Amino Acid & Score \\
    \hline
    Isoleucine (I)        &  4.5  \\
    Valine (V)            &  4.2  \\
    Leucine (L)           &  3.8  \\
    Phenylalanine (F)     &  2.8  \\
    Cysteine (C)          &  2.5  \\
    Methionine (M)        &  1.9  \\
    Alanine (A)           &  1.8  \\
    Glycine (G)           & –0.4  \\
    Threonine (T)         & –0.7  \\
    Serine (S)            & –0.8  \\
    Tryptophan (W)        & –0.9  \\
    Tyrosine (Y)          & –1.3  \\
    Proline (P)           & –1.6  \\
    Histidine (H)         & –3.2  \\
    Asparagine (N)        & –3.5  \\
    Aspartic acid (D)     & –3.5  \\
    Glutamine (Q)         & –3.5  \\
    Glutamic acid (E)     & –3.5  \\
    Lysine (K)            & –3.9  \\
    Arginine (R)          & –4.5  \\
    \hline
  \end{tabular}
\end{table}

The Kyte–Doolittle hydrophobicity scale\cite{KyteDoolittle1982} is a standard method for locating hydrophobic segments in proteins. Positive scores indicate hydrophobic regions. By averaging these scores over a sliding window, one can tune the analysis to different structural features: a window of 5–7 residues is typically best for spotting surface‐exposed patches, whereas a window of 19–21 residues highlights transmembrane helices when the average exceeds roughly 1.6. These thresholds are useful guidelines, but exceptions do occur. \texttt{BEER} implements the Kyte-Doolittle hydrophobicity scale, computing hydrophobicity via sliding-window averages (Eq.~\ref{Eq:KD})

\begin{equation}
    H(i) = \frac{1}{W}\sum_{j=i}^{i+W-1} H_j
    \label{Eq:KD}
\end{equation}

where $H_j$ is the Kyte-Doolittle score for residue $j$, and $W$ is the window size. The hydrophobicity scores ($H_j$) of individual amino acids given by the Kyte-Dolittle scale are presented in Table~\ref{tab:KD}.

\subsection{Net Charge and Isoelectric Point}
At any specified \emph{pH}, the charge state of ionisable groups in a protein can be approximated from the Henderson–Hasselbalch (H–H) relation (Eq.~\ref{eq:hh}).

\begin{equation}
    \label{eq:hh}
    \mathrm{p}K_a - \mathrm{pH} = \log_{10}\!\left(\frac{[\mathrm{HA}]}{[\mathrm{A}^{-}]}\right)
\end{equation}
where $[\mathrm{HA}]$ and $[\mathrm{A}^{-}]$ are the molar concentrations of the protonated and deprotonated forms of a generic weak acid. 

Re-arranging Eq.~\ref{eq:hh} gives the fractional \emph{de}protonation for each ionisable site:
\begin{align}
    f_{\text{prot}} &= \frac{1}{1 + 10^{\mathrm{pH}-pK_a}} \label{eq:fprot} \\
    f_{\text{deprot}} &= 1 - f_{\text{prot}}            \label{eq:fdeprot}
\end{align}

The net charge $Q_{\text{net}}(\mathrm{pH})$ of a protein is obtained by summing the contributions from  

\begin{itemize}
  \item the N-terminal $\alpha$–amino group,  
  \item the C-terminal $\alpha$–carboxyl group, and  
  \item all side-chain acids or bases in the sequence.
\end{itemize}

For basic sites (N-terminus, Lys, Arg, His) the protonated state carries a $+1$ charge; for acidic sites (C-terminus, Asp, Glu, Cys, Tyr) the deprotonated state carries a $-1$ charge. Hence, $Q_{\text{net}}(\mathrm{pH})$ can be given by Eq. \ref{eq:qnet}.

\begin{equation}
  \label{eq:qnet}
  Q_{\text{net}}(\mathrm{pH}) = 
      \underbrace{f_{\text{prot}}^{\text{N-term}}}_{+\!1}
    - \underbrace{f_{\text{deprot}}^{\text{C-term}}}_{-\!1}
    + \sum_{i\in\mathcal{B}}   f_{\text{prot}}^{(i)}
    - \sum_{j\in\mathcal{A}}   f_{\text{deprot}}^{(j)},
\end{equation}
where $\mathcal{B}=\{\mathrm{K,R,H}\}$ and $\mathcal{A}=\{\mathrm{D,E,C,Y}\}$ denote the basic and acidic residue sets, respectively.

\texttt{BEER} adopts the single-site \emph{p}K\textsubscript{a} constants recommended by \texttt{EMBOSS}/\texttt{BioPython} (Table~\ref{tab:pka-values}). These values ultimately trace back to Ref.~\cite{Bjellqvist1993} with minor refinements. They provide good agreement with measured average pI values for soluble proteins.

\begin{table}[ht]
  \centering
  \caption{Ionisable groups, their charges in the two protonation states, and the
           single–site $pK_a$ values used by \textsc{BEER}.}
  \label{tab:pka-values}
  \begin{tabular}{lccccc}
    \hline
    \multirow{2}{*}{\textbf{Group}} & \multirow{2}{*}{\textbf{Symbol}} &
    \multicolumn{2}{c}{\textbf{Net charge}} & \multirow{2}{*}{$\boldsymbol{pK_a}$} \\
    \cline{3-4}
    & & protonated & de-protonated & \\[2pt]
    \hline
    N-terminus              & ---NH$_3^{+}$ & $+1$ & $0$  & 9.69 \\
    C-terminus              & ---COOH       & $0$  & $-1$ & 2.34 \\
    Aspartate               & D             & $0$  & $-1$ & 3.90 \\
    Glutamate               & E             & $0$  & $-1$ & 4.07 \\
    Cysteine                & C             & $0$  & $-1$ & 8.18 \\
    Tyrosine                & Y             & $0$  & $-1$ & 10.46 \\
    Histidine               & H             & $+1$ & $0$  & 6.04 \\
    Lysine                  & K             & $+1$ & $0$  & 10.54 \\
    Arginine                & R             & $+1$ & $0$  & 12.48 \\
    \hline
  \end{tabular}
\end{table}

These constants assume an aqueous environment with minimal site–site coupling. Although local microenvironments and neighbouring charges will shift the true microscopic \emph{p}K\textsubscript{a} values, Eq.~\eqref{eq:qnet} with the above parameters typically predicts the experimental isoelectric point (pI) of folded proteins within $\pm0.3$\,pH units\cite{Sillero1989,Bjellqvist1993}—sufficient for formulation screening and surface-patch mapping.

A sliding‐window net-charge curve is produced by evaluating Eq.~\eqref{eq:qnet} for \mbox{pH = 0\,–\,14} in 0.1-unit steps. To determine the pI, \texttt{BEER} locates the pH where $Q_{\text{net}}(\mathrm{pH})=0$ by a one-dimensional root search (bisection) over the same range. The routine is vectorised in \texttt{NumPy} for speed and returns both the full charge curve and the scalar pI. Users may override the default \emph{p}H and \emph{p}K\textsubscript{a} values in the \texttt{Settings} tab.

In day-to-day work, the charge curve offers immediate guidance on buffer selection: regions where $|Q_{\text{net}}|$ is small are sensitive to small pH shifts and therefore prone to aggregation or precipitation. Combined with the Kyte–Doolittle hydrophobicity profile, researchers can rapidly spot amphipathic helices or acidic/basic clusters, informing construct design, mutagenesis, or fusion-tag placement.

\subsection{Physicochemical Properties}

\textsc{BEER} reports five widely used physicochemical descriptors, each calculated with the canonical formulae implemented in \texttt{BioPython}. All values are displayed in the \emph{Properties} tab and exported with the PDF or text report.

\subsubsection{Molecular weight (MW)}
The average isotopic mass is obtained by summing the residue masses of all amino acids and subtracting $\left(n_{\text{res}}-1\right)$ water molecules to account for peptide-bond formation (Eq.~\ref{eq:mw}).
\begin{equation}
  MW = \sum_{i=1}^{n_{\text{res}}} m_i \;-\;(n_{\text{res}}-1)\,18.015~\text{Da}.
  \label{eq:mw}
\end{equation}

The individual residue masses ($m_i$) follow the IUPAC 1998 average atomic-mass table.

\subsubsection{Extinction coefficient at 280\,nm}  
Protein absorbance in the near-UV is dominated by \textit{Trp}, \textit{Tyr}, and disulfide-linked \textit{Cystine}. \texttt{BEER} uses the Gill--von Hippel empirical relation\cite{GillVonHippel1989} (Eq.~\ref{eq:GvH}) assuming all cysteines are oxidized unless the user selects “\textit{Assume reducing conditions}” in \texttt{Settings}.

\begin{align}
  \varepsilon_{280}
  &= 5500\,N_{\mathrm{Trp}} + 1490\,N_{\mathrm{Tyr}} \notag\\
  &\quad+ 125\,N_{\mathrm{Cystine}}\quad(\mathrm{M}^{-1}\,\mathrm{cm}^{-1})
    \label{eq:GvH}
\end{align}

\subsubsection{GRAVY score}  
The Grand Average of Hydropathicity (GRAVY) is the mean Kyte--Doolittle value over the entire sequence (Eq.~\ref{eq:gravy}).
\begin{equation}    
    \text{GRAVY} = \langle H\rangle = \frac{1}{n_{\text{res}}}\sum_{i=1}^{n_{\text{res}}} H_i,
    \label{eq:gravy}
\end{equation}

where $H_i$ is the hydropathy index of residue $i$. Positive values indicate overall hydrophobic character.

\subsubsection{Solubility prediction}
\texttt{BEER} gives a solubility prediction based on the GRAVY score. Large expression datasets indicate that proteins with $\langle H\rangle\lesssim0.5$ are routinely recovered in the soluble fraction, whereas more hydrophobic chains tend to aggregate or partition into membranes. BEER therefore applies the decision rule  
$\text{if }\langle H\rangle<0.5\;\Rightarrow\;\text{``likely soluble''},  \quad \text{else }\;\text{``low solubility predicted''}.$

The threshold is rescaled automatically if the user changes the hydropathy sliding-window size, allowing local surface patches to be emphasised instead of the global average.  Although the heuristic of these equations is deliberately simple, comparative tests against empirical solubility data for \textit{E.\,coli} and insect-cell expression systems show accuracies on par with more elaborate predictors (\citealp{KyteDoolittle1982,Bjellqvist1993,Sillero1989}).  Consequently, it offers an instantaneous \textit{first-pass} screen before committing resources to detailed solubility optimization.

\subsubsection{Aromaticity}  
Aromaticity is defined as the fraction of aromatic residues (Eq.~\ref{eq:aro}).

\begin{equation}    
    \text{Aromaticity} = \frac{N_{\text{Phe}} + N_{\text{Trp}}
                             + N_{\text{Tyr}}}
                            {n_{\text{res}}}
    \label{eq:aro}    
\end{equation}

this metric correlates with UV absorbance and, to some extent, core packing density.

\subsubsection{Instability index}  
Following Guruprasad \textit{et al.}\ \cite{Guruprasad1990}, every dipeptide is assigned an instability weight \(I_{jk}\).  The index
is the average weight scaled to 100 (Eq.~\ref{eq:instab}).

\begin{equation}
  \Pi = \frac{10}{n_{\text{res}}}\,
             \sum_{i=1}^{n_{\text{res}}-1} I_{\,\text{res}_i\,\text{res}_{i+1}}.
    \label{eq:instab}
\end{equation}

Empirically, proteins with $\Pi < 40$ are considered \emph{stable} in solution, whereas those with $\Pi > 40$ are prone to degradation or aggregation.

Together, these five descriptors provide a rapid overview of size, optical detectability, hydrophobic balance, aromatic content, and expected stability, helping users decide on expression systems, buffer conditions, and further modeling steps.

\section{Examples}

\begin{figure}
    \centering
    \includegraphics[width=1\linewidth]{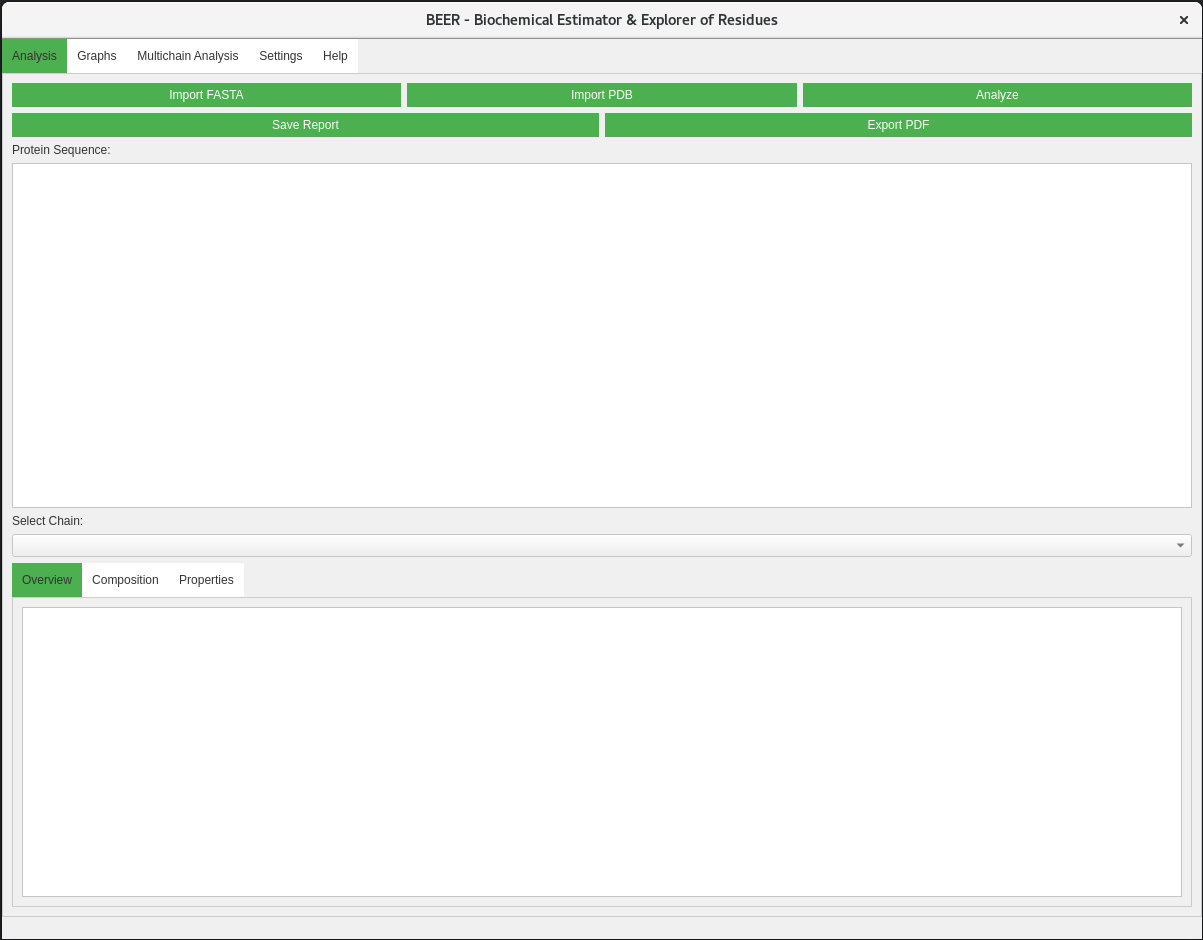}
    \caption{Main \texttt{BEER} window offering a clean and easy to navigate Graphical User Interface (GUI).}
    \label{fig:main}
\end{figure}

The \texttt{BEER} software offers a GUI that is easy to use and navigate. Fig.~\ref{fig:main} shows the main window that open on executing the code. Both \textsl{light} and \textsl{dark} themes are available that can be toggled in the \texttt{settings} tab.

The \texttt{settings} tab offers several customizable options, which will be mentioned in the following discussion. To demonstrate \texttt{BEER} in practice, we present two worked examples. 

\subsection{Example 1. Single-chain inspection}
\subsubsection{Analysis}
\begin{figure}
    \centering
    \includegraphics[width=1\linewidth]{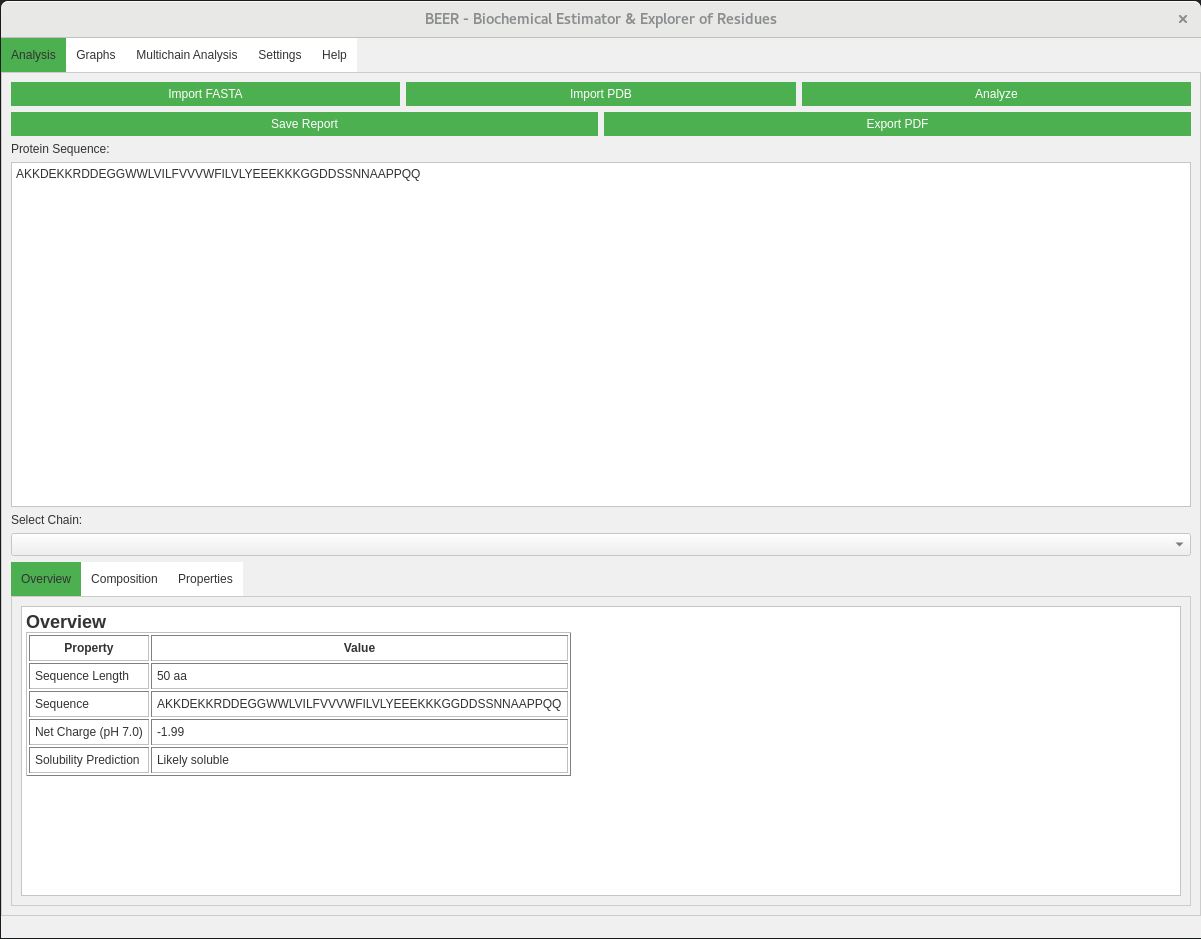}
    \caption{Main \texttt{BEER} window after analyzing the 50-residue test peptide. Tabs for overview, composition and properties are stacked above interactive plots.}
    \label{fig:overviewGUI}
\end{figure}

The user has the option to import a \texttt{FASTA} sequence file or a \texttt{PDB} file through the \texttt{Import FASTA} or \texttt{Import PDB} buttons on the top. Alternatively, the GUI also offers a editable text field where a sequence (1-lettered) can be pasted or typed in. For example, the 50‐residue construct \textsl{AKKDEKKRDDEGGWWLVILFVVVWFILVLYEEEKKKGGDDSSNNAAPPQQ} was pasted directly into the sequence field and analyzed with default settings (window~$W=9$, pH $=7.0$).  Fig.~\ref{fig:overviewGUI} shows the GUI immediately after pressing \texttt{Analyze}. It is to be noted that the window and pH values can be changed in \texttt{settings}, after which the \texttt{Apply Settings} and the \texttt{Analyze} (in the main \texttt{Analysis} tab) must be pressed.

The \texttt{Overview} tab in the bottom half of the GUI shows the sequence length, net charge at pH 7.0 (also at other user selected pH) and predicted solubility, along with a reproduction of the entered sequence. The \texttt{Composition} tab displays a table of amino acid composition with three columns: a) 1-letter amino acid names, b) Number of each amino acid (Count), and c) its \% frequency. The table can be sorted on the fly with four buttons:
\begin{itemize}
  \item alphabetical order (\texttt{A\,\textrightarrow\,Z});
  \item by decreasing percentage (\texttt{By~Comp});
  \item by increasing hydrophobicity (\texttt{Hydro~$\uparrow$});
  \item by decreasing hydrophobicity (\texttt{Hydro~$\downarrow$}).
\end{itemize}

\begin{figure}
    \centering
    \includegraphics[width=1\linewidth]{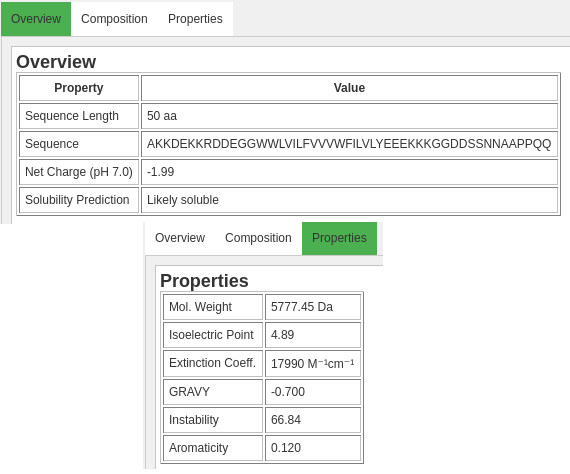}
    \caption{Spashots showing the outputs in the \texttt{Overview} and \texttt{Properties} tabs.}
    \label{fig:tabsnaps}
\end{figure}

By default, the table is sorted by composition. The other information are displayed in the \texttt{Properties} tab. These include molecular weight of the protein chain, isoelectric point, extinction coefficient at 280 nm, GRAVY score, instability index, and aromaticity. Snapshots of the \texttt{Overview} and \texttt{Properties} tabs for the present protein construct are shown in Fig.~\ref{fig:tabsnaps}. Due to the length of the \texttt{Composition} tab, it is not shown here.

\subsubsection{Graphs}
The \texttt{graphs} tab on the top of the GUI shows seven plots related to the sequence analyzed.

\paragraph*{Composition.}

\begin{figure}[b]
    \centering
    \includegraphics[width=1\linewidth]{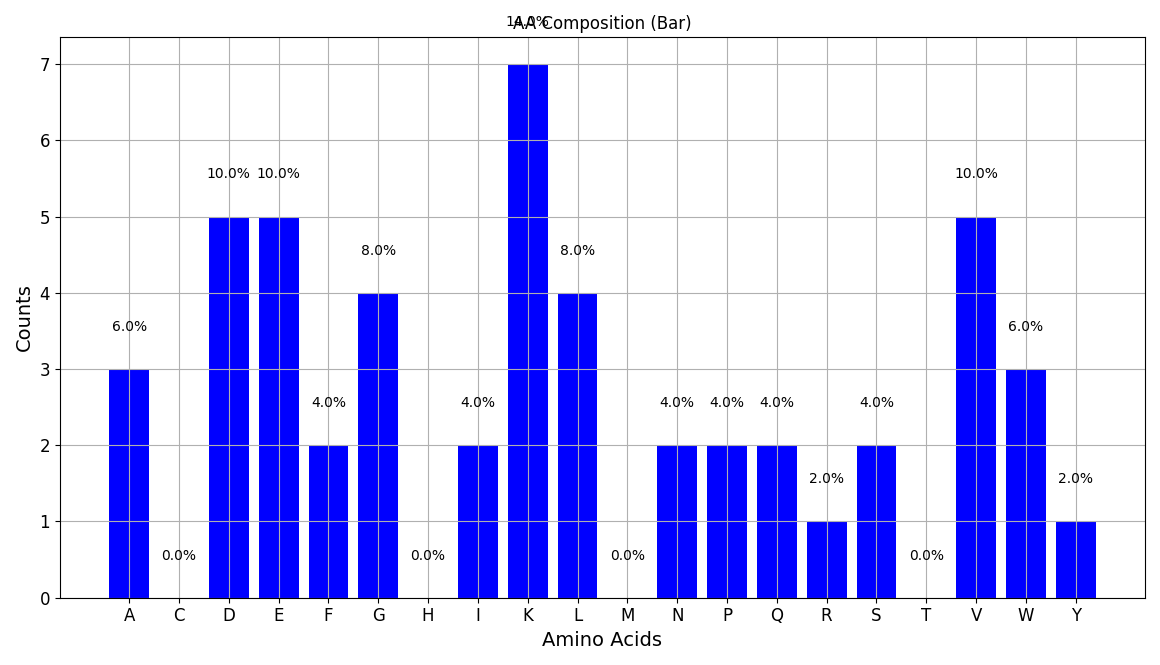}
    \caption{Amino acid composition displayed a bar diagram.}
    \label{fig:compBar}
\end{figure}

\begin{figure}[t]
    \centering
    \includegraphics[width=1\linewidth]{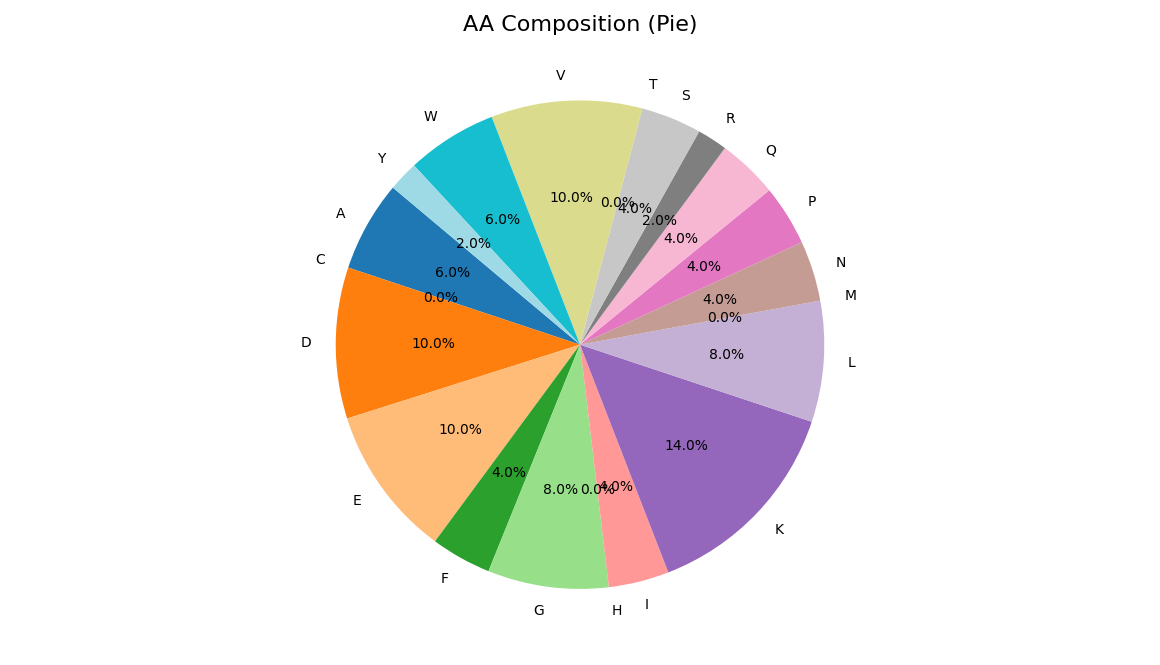}
    \caption{Amino acid composition displayed a pie diagram.}
    \label{fig:compPie}
\end{figure}

Fig.~\ref{fig:compBar} and Fig.~\ref{fig:compPie} show the amino acid composition of the chain as a bar and a pie plot respectively. They reveal three enriched residue classes: (i) charged Lys/Asp/Glu (36\,\%), (ii) hydrophobic Leu/Val (20\,\%), and (iii) Gly (12\,\%).

These plots give an instant impression of the chemical nature of the sequence being analyzed.

\begin{figure}[h]
    \centering
    \includegraphics[width=1\linewidth]{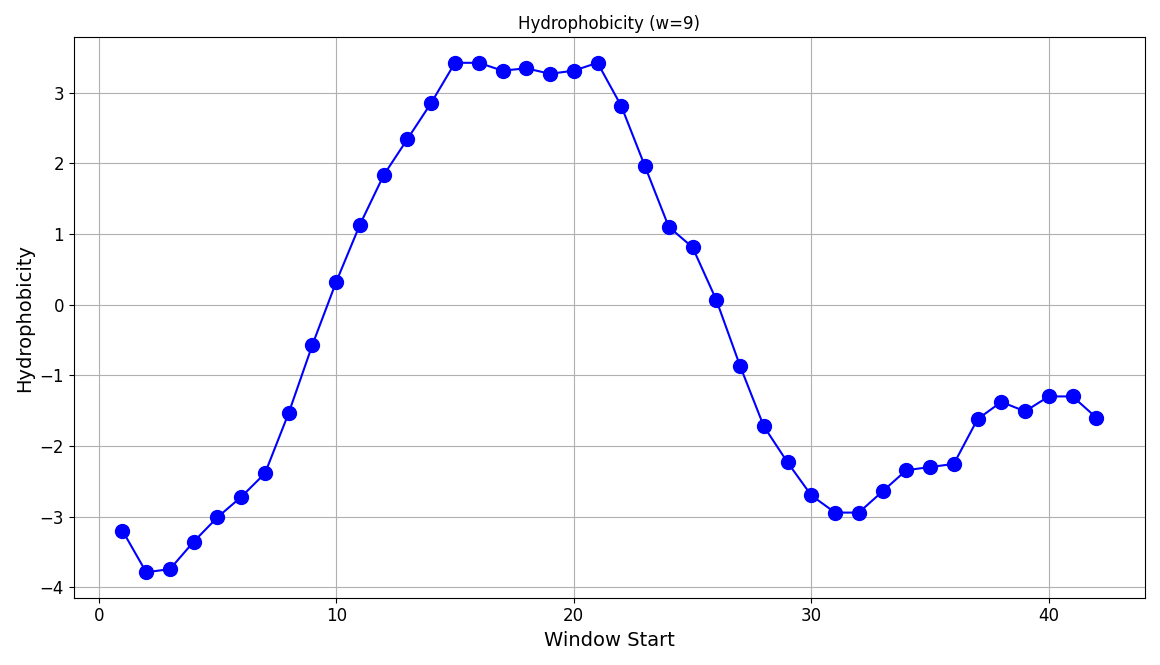}
    \caption{Kyte–Doolittle hydrophobicity profile (window $W=9$) for the given sequence.}
    \label{fig:hydroCurve}
\end{figure}

\paragraph*{Hydrophobicity profile.}
The sliding‐window hydrophobicity curve (Fig.~\ref{fig:hydroCurve}) contains a pronounced peak spanning residues~15–30 (window average $H\!>\!2.0$).  This shows a contiguous hydrophobic patch in the sequence long enough to form a membrane helix or aggregation nucleus.

\paragraph*{Charge plot and pI.}

\begin{figure}[b]
    \centering
    \includegraphics[width=1\linewidth]{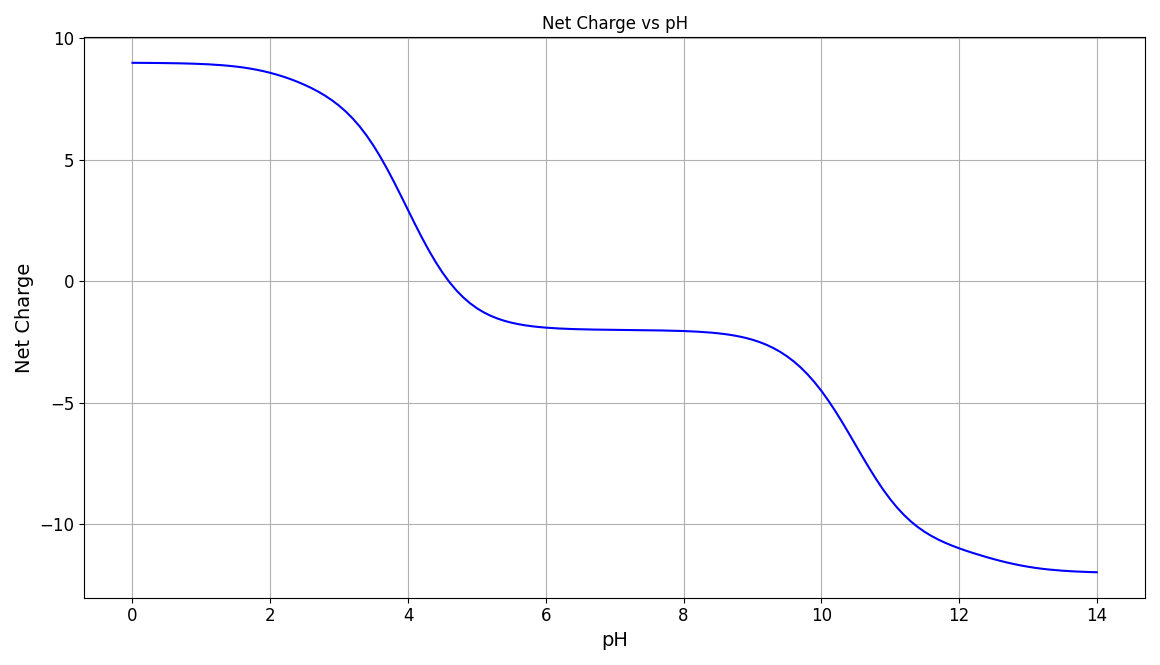}
    \caption{Net charge as a function of pH for the single-chain construct.}
    \label{fig:pH}
\end{figure}

The pH–charge graph (Fig.~\ref{fig:pH}) crosses zero net charge at $\text{pH}=4.89$, which is the $\text{pI}$ for this protein construct. At physiological pH (7.4) the net charge is $-2.08$, arising mainly from the Asp/Glu cluster N‐terminal to the hydrophobic stretch. Such a distribution suggests that the hydrophobic core could be solubilized by the surrounding acidic shell—a typical feature of intrinsically disordered, amphipathic peptides.

\begin{figure}[h]
    \centering
    \includegraphics[width=1\linewidth]{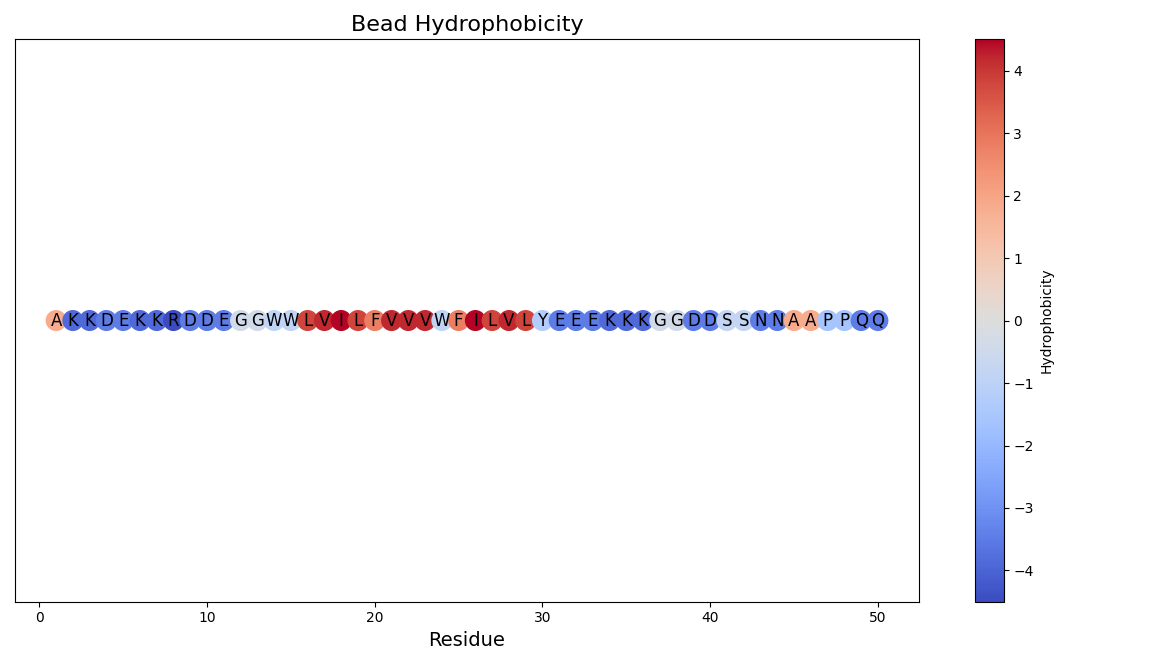}
    \caption{Bead model showing Kyte-Dolittle hydrophobicty given by the color bar.}
    \label{fig:beadHydro}
\end{figure}

\paragraph*{Bead‐model visualizations.}
Figure~\ref{fig:beadHydro} shows the per‐residue hydrophobicity bead model: each residue is represented as a circle whose color reflects its Kyte–Doolittle score (blue = hydrophilic, red = hydrophobic). The contiguous run of red beads between positions 15 and 30 highlights the strongly hydrophobic core of this peptide, suggesting a potential membrane‐inserting or aggregation‐prone segment.  

\begin{figure}[h]
    \centering
    \includegraphics[width=1\linewidth]{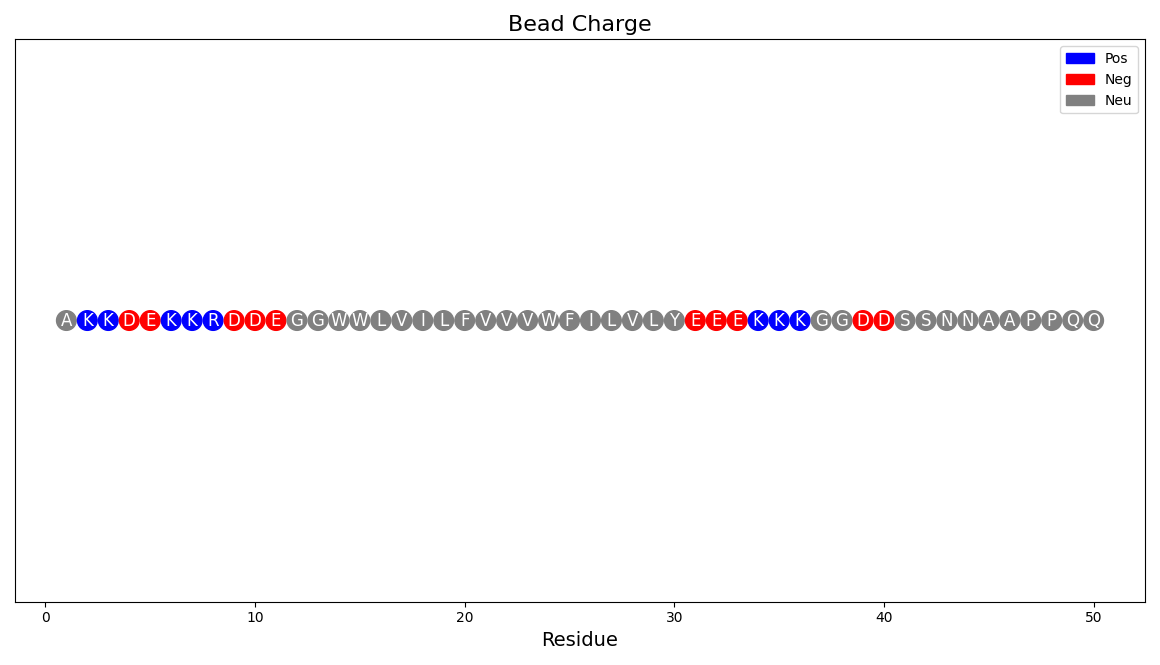}
    \caption{Bead model showing amino acid charges.}
    \label{fig:beadCharge}
\end{figure}

Figure~\ref{fig:beadCharge} presents the charge bead model: positively charged residues (Lys, Arg, His) are shown in blue, negatively charged residues (Asp, Glu) in red, and neutral residues in gray. 

\begin{figure}
    \centering
    \includegraphics[width=1\linewidth]{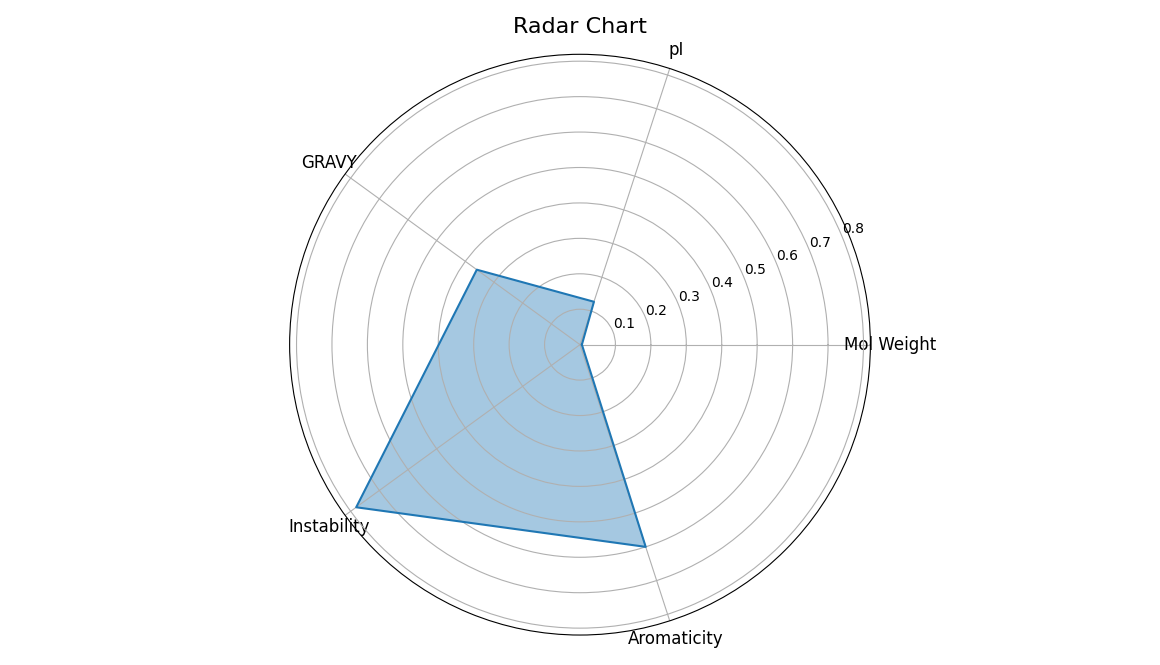}
    \caption{Radar chart summarizing key physicochemical properties on a normalized scale.}
    \label{fig:radarChart}
\end{figure}

\paragraph*{Properties radar chart.}
Figure~\ref{fig:radarChart} overlays five key physicochemical properties—molecular weight, isoelectric point, GRAVY score, instability index, and aromaticity—on a normalized radar plot. Together, these features provide a concise “fingerprint” of the peptide’s global biophysical profile.

The grids, figure headings, and bead labels (for bead plots) can be toggled on/off in the \texttt{settings} tab. The axis and tick font sizes, graph colors, colormaps (for hydrophobicity bead plot) are also customizable. The graphs can be saved in PNG, PDF, or SVG formats (as selected in \texttt{settings}) individually or all at once. The interactive nature of these plot windows allows the user a number of customizations such as zooming in on certain regions, identifying the coordinates of a point by clicking/hovering on it, and moving the plot for proper visualization. \texttt{Save Report} and \texttt{Export PDF} produces a multipage report (respectively in TXT and PDF formats) collating the chain’s properties and  plots, ideal for lab records and supplementary materials.

\subsection{Example 2. Multichain comparison}

\begin{figure}[b]
    \centering
    \includegraphics[width=1\linewidth]{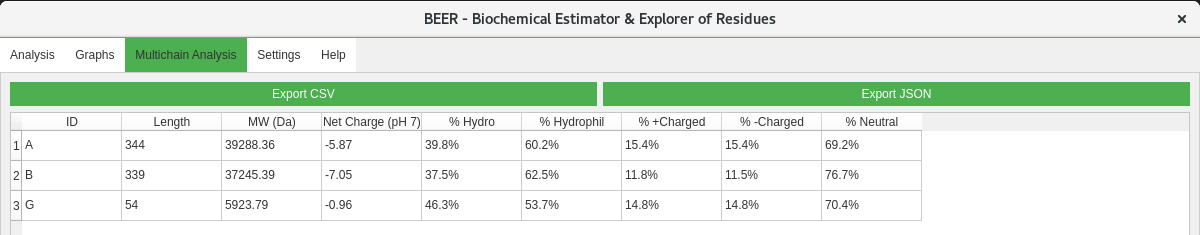}
    \caption{The multichain comparison table for \texttt{1GP2}.}
    \label{fig:multi}
\end{figure}

A trimeric assembly (\texttt{PDB} ID: 1GP2) containing three distinct chains (A, B, and G) was loaded via \texttt{Import PDB}. \texttt{BEER} automatically extracted each sequence and populated the multichain summary table (Fig.~\ref{fig:multi}).

\paragraph*{Rapid screening.}
Chain A (344 residues, 39.3 kDa) carries a net charge of \(-5.87\) at pH 7.0 with 39.8 \% hydrophobic residues. Chain B (339 residues, 37.2 kDa) is slightly shorter and more acidic (net charge \(-7.05\)) with 37.5 \% hydrophobic content. In contrast, Chain G is a 54-residue peptide (5.9 kDa) with the highest hydrophobic fraction (46.3 \%) and a near-neutral net charge (\(-0.96\)). These metrics flag Chain G as unusually hydrophobic for its size, suggesting a compact amphipathic segment.

\paragraph*{Interactive chain selection.}
In the main \texttt{Analysis} tabs, any of the available chains can be selected and the single-chain analysis can be performed on it. This allows comparison of the chains visually with the help of the graphs. For example, inspecting Chain G reveals a pronounced hydrophobic peak between residues 20–30 and a balanced charge profile, consistent with a membrane-interacting helix flanked by modestly charged termini.

\paragraph*{Export.}
Selecting \texttt{Export CSV} or \texttt{Export JSON} saves the complete multichain summary (Fig.~\ref{fig:multi}) in the corresponding format.

\section{Conclusion}

\texttt{BEER} offers a streamlined, user-friendly desktop GUI for rapid physicochemical profiling of protein sequences. It computes amino acid composition, sliding-window hydrophobicity, pH-dependent net charge with automatic pI determination, solubility heuristics, and key indices (molecular weight, 280 nm extinction coefficient, GRAVY, instability, and aromaticity). Interactive visualizations—including bar and pie charts, bead models, and radar diagrams—combined with multichain comparison empower analysis of sequences up to 10\,000 residues in under one second on standard hardware.  

Nevertheless, \texttt{BEER} has limitations: the GRAVY-based solubility heuristic provides only a first-pass estimate and may not capture context-dependent aggregation behavior; the single-site Henderson–Hasselbalch model, while user-overridable, does not account for microenvironment coupling that can shift true $pK_a$ values; noncanonical amino acids, post-translational modifications, and nucleic acid components are currently unsupported; and structural analysis—such as secondary-structure annotation or solvent-accessibility mapping—is not yet included. 

Looking ahead, we plan to add structural modules for overlaying computed metrics on three-dimensional models via NGLview, and to extend sequence support to include nonstandard residues and common modifications. We will also implement automated database queries to UniProt and PDB, expose a scripting API with plugin capability for community-driven extensions, and offer customizable report templates (HTML, Word, LaTeX) alongside potential laboratory information management system integration. These developments will further bridge rapid sequence-based insights and detailed experimental or structural workflows, making \texttt{BEER} an even more powerful tool for teaching and research alike.  

\section{Availability}

\texttt{BEER} is distributed under the GNU General Public License v3 and is available on GitHub at \url{https://github.com/chemgame/BEER}. The software has been tested on CentOS Stream 8, macOS Sequoia 15.3.2, and Windows 11. We recommend installation in a dedicated Conda environment with the following packages: \texttt{biopython}, \texttt{matplotlib}, \texttt{PyQt5}, and \texttt{mplcursors}.

\bibliographystyle{apsrev4-2}
\bibliography{ref}

\begin{thebibliography}{9}%
\makeatletter
\providecommand \@ifxundefined [1]{%
 \@ifx{#1\undefined}
}%
\providecommand \@ifnum [1]{%
 \ifnum #1\expandafter \@firstoftwo
 \else \expandafter \@secondoftwo
 \fi
}%
\providecommand \@ifx [1]{%
 \ifx #1\expandafter \@firstoftwo
 \else \expandafter \@secondoftwo
 \fi
}%
\providecommand \natexlab [1]{#1}%
\providecommand \enquote  [1]{``#1''}%
\providecommand \bibnamefont  [1]{#1}%
\providecommand \bibfnamefont [1]{#1}%
\providecommand \citenamefont [1]{#1}%
\providecommand \href@noop [0]{\@secondoftwo}%
\providecommand \href [0]{\begingroup \@sanitize@url \@href}%
\providecommand \@href[1]{\@@startlink{#1}\@@href}%
\providecommand \@@href[1]{\endgroup#1\@@endlink}%
\providecommand \@sanitize@url [0]{\catcode `\\12\catcode `\$12\catcode
  `\&12\catcode `\#12\catcode `\^12\catcode `\_12\catcode `\%12\relax}%
\providecommand \@@startlink[1]{}%
\providecommand \@@endlink[0]{}%
\providecommand \url  [0]{\begingroup\@sanitize@url \@url }%
\providecommand \@url [1]{\endgroup\@href {#1}{\urlprefix }}%
\providecommand \urlprefix  [0]{URL }%
\providecommand \Eprint [0]{\href }%
\providecommand \doibase [0]{https://doi.org/}%
\providecommand \selectlanguage [0]{\@gobble}%
\providecommand \bibinfo  [0]{\@secondoftwo}%
\providecommand \bibfield  [0]{\@secondoftwo}%
\providecommand \translation [1]{[#1]}%
\providecommand \BibitemOpen [0]{}%
\providecommand \bibitemStop [0]{}%
\providecommand \bibitemNoStop [0]{.\EOS\space}%
\providecommand \EOS [0]{\spacefactor3000\relax}%
\providecommand \BibitemShut  [1]{\csname bibitem#1\endcsname}%
\let\auto@bib@innerbib\@empty
\bibitem [{\citenamefont {Kyte}\ and\ \citenamefont
  {Doolittle}(1982)}]{KyteDoolittle1982}%
  \BibitemOpen
  \bibfield  {author} {\bibinfo {author} {\bibfnamefont {J.}~\bibnamefont
  {Kyte}}\ and\ \bibinfo {author} {\bibfnamefont {R.~F.}\ \bibnamefont
  {Doolittle}},\ }\href {https://doi.org/10.1016/0022-2836(82)90515-0}
  {\bibfield  {journal} {\bibinfo  {journal} {Journal of Molecular Biology}\
  }\textbf {\bibinfo {volume} {157}},\ \bibinfo {pages} {105} (\bibinfo {year}
  {1982})}\BibitemShut {NoStop}%
\bibitem [{\citenamefont {Henderson}(1908)}]{Henderson1908}%
  \BibitemOpen
  \bibfield  {author} {\bibinfo {author} {\bibfnamefont {L.~J.}\ \bibnamefont
  {Henderson}},\ }\href@noop {} {\bibfield  {journal} {\bibinfo  {journal}
  {American Journal of Physiology}\ }\textbf {\bibinfo {volume} {21}},\
  \bibinfo {pages} {173} (\bibinfo {year} {1908})}\BibitemShut {NoStop}%
\bibitem [{\citenamefont {Guruprasad}\ \emph {et~al.}(1990)\citenamefont
  {Guruprasad}, \citenamefont {Reddy},\ and\ \citenamefont
  {Pandit}}]{Guruprasad1990}%
  \BibitemOpen
  \bibfield  {author} {\bibinfo {author} {\bibfnamefont {K.}~\bibnamefont
  {Guruprasad}}, \bibinfo {author} {\bibfnamefont {B.~V.~B.}\ \bibnamefont
  {Reddy}},\ and\ \bibinfo {author} {\bibfnamefont {M.~W.}\ \bibnamefont
  {Pandit}},\ }\href {https://doi.org/10.1093/protein/4.2.155} {\bibfield
  {journal} {\bibinfo  {journal} {Protein Engineering, Design and Selection}\
  }\textbf {\bibinfo {volume} {4}},\ \bibinfo {pages} {155} (\bibinfo {year}
  {1990})}\BibitemShut {NoStop}%
\bibitem [{\citenamefont {Cock}\ \emph {et~al.}(2009)\citenamefont {Cock},
  \citenamefont {Antao}, \citenamefont {Chang}, \citenamefont {Chapman},
  \citenamefont {Cox}, \citenamefont {Dalke}, \citenamefont {Friedberg},
  \citenamefont {Hamelryck}, \citenamefont {Kauff}, \citenamefont
  {Wilczynski},\ and\ \citenamefont {de~Hoon}}]{Cock2009}%
  \BibitemOpen
  \bibfield  {author} {\bibinfo {author} {\bibfnamefont {P.~J.~A.}\
  \bibnamefont {Cock}}, \bibinfo {author} {\bibfnamefont {T.}~\bibnamefont
  {Antao}}, \bibinfo {author} {\bibfnamefont {J.~T.}\ \bibnamefont {Chang}},
  \bibinfo {author} {\bibfnamefont {B.~A.}\ \bibnamefont {Chapman}}, \bibinfo
  {author} {\bibfnamefont {C.~J.}\ \bibnamefont {Cox}}, \bibinfo {author}
  {\bibfnamefont {A.}~\bibnamefont {Dalke}}, \bibinfo {author} {\bibfnamefont
  {I.}~\bibnamefont {Friedberg}}, \bibinfo {author} {\bibfnamefont
  {T.}~\bibnamefont {Hamelryck}}, \bibinfo {author} {\bibfnamefont
  {F.}~\bibnamefont {Kauff}}, \bibinfo {author} {\bibfnamefont
  {B.}~\bibnamefont {Wilczynski}},\ and\ \bibinfo {author} {\bibfnamefont
  {M.~J.~L.}\ \bibnamefont {de~Hoon}},\ }\href
  {https://doi.org/10.1093/bioinformatics/btp163} {\bibfield  {journal}
  {\bibinfo  {journal} {Bioinformatics}\ }\textbf {\bibinfo {volume} {25}},\
  \bibinfo {pages} {1422} (\bibinfo {year} {2009})}\BibitemShut {NoStop}%
\bibitem [{\citenamefont {{Riverbank Computing Limited}}(2025)}]{PyQt5}%
  \BibitemOpen
  \bibfield  {author} {\bibinfo {author} {\bibnamefont {{Riverbank Computing
  Limited}}},\ }\href@noop {} {\bibinfo {title} {{PyQt5}: {Python} bindings for
  the qt application framework}},\ \bibinfo {howpublished}
  {\url{https://www.riverbankcomputing.com/software/pyqt/}} (\bibinfo {year}
  {2025}),\ \bibinfo {note} {accessed 23~Apr~2025}\BibitemShut {NoStop}%
\bibitem [{\citenamefont {Hunter}(2007)}]{Hunter2007}%
  \BibitemOpen
  \bibfield  {author} {\bibinfo {author} {\bibfnamefont {J.~D.}\ \bibnamefont
  {Hunter}},\ }\href {https://doi.org/10.1109/MCSE.2007.55} {\bibfield
  {journal} {\bibinfo  {journal} {Computing in Science \& Engineering}\
  }\textbf {\bibinfo {volume} {9}},\ \bibinfo {pages} {90} (\bibinfo {year}
  {2007})}\BibitemShut {NoStop}%
\bibitem [{\citenamefont {Bjellqvist}\ \emph {et~al.}(1993)\citenamefont
  {Bjellqvist}, \citenamefont {Hughes}, \citenamefont {Pasquali}, \citenamefont
  {Paquet}, \citenamefont {Ravier}, \citenamefont {Sanchez}, \citenamefont
  {Frutiger},\ and\ \citenamefont {Hochstrasser}}]{Bjellqvist1993}%
  \BibitemOpen
  \bibfield  {author} {\bibinfo {author} {\bibfnamefont {B.}~\bibnamefont
  {Bjellqvist}}, \bibinfo {author} {\bibfnamefont {G.~J.}\ \bibnamefont
  {Hughes}}, \bibinfo {author} {\bibfnamefont {C.}~\bibnamefont {Pasquali}},
  \bibinfo {author} {\bibfnamefont {N.}~\bibnamefont {Paquet}}, \bibinfo
  {author} {\bibfnamefont {F.}~\bibnamefont {Ravier}}, \bibinfo {author}
  {\bibfnamefont {J.~C.}\ \bibnamefont {Sanchez}}, \bibinfo {author}
  {\bibfnamefont {S.}~\bibnamefont {Frutiger}},\ and\ \bibinfo {author}
  {\bibfnamefont {D.}~\bibnamefont {Hochstrasser}},\ }\href
  {https://doi.org/10.1002/elps.11501401163} {\bibfield  {journal} {\bibinfo
  {journal} {Electrophoresis}\ }\textbf {\bibinfo {volume} {14}},\ \bibinfo
  {pages} {1023} (\bibinfo {year} {1993})}\BibitemShut {NoStop}%
\bibitem [{\citenamefont {Sillero}\ and\ \citenamefont
  {Sillero}(1989)}]{Sillero1989}%
  \BibitemOpen
  \bibfield  {author} {\bibinfo {author} {\bibfnamefont {A.}~\bibnamefont
  {Sillero}}\ and\ \bibinfo {author} {\bibfnamefont {I.}~\bibnamefont
  {Sillero}},\ }\href {https://doi.org/10.1016/0307-4412(89)90069-8} {\bibfield
   {journal} {\bibinfo  {journal} {Biochemical Education}\ }\textbf {\bibinfo
  {volume} {17}},\ \bibinfo {pages} {158} (\bibinfo {year} {1989})}\BibitemShut
  {NoStop}%
\bibitem [{\citenamefont {Gill}\ and\ \citenamefont {von
  Hippel}(1989)}]{GillVonHippel1989}%
  \BibitemOpen
  \bibfield  {author} {\bibinfo {author} {\bibfnamefont {S.~C.}\ \bibnamefont
  {Gill}}\ and\ \bibinfo {author} {\bibfnamefont {P.~H.}\ \bibnamefont {von
  Hippel}},\ }\href {https://doi.org/10.1016/0003-2697(89)90602-7} {\bibfield
  {journal} {\bibinfo  {journal} {Analytical Biochemistry}\ }\textbf {\bibinfo
  {volume} {182}},\ \bibinfo {pages} {319} (\bibinfo {year}
  {1989})}\BibitemShut {NoStop}%
\end{thebibliography}%

\end{document}